# High power highly stable passively Q-switched fiber laser based on monolayer graphene

**Hanshuo Wu[1], Jiaxin Song[1], Jian Wu[1,2,*], Jiangming Xu[1,2], Hu Xiao[1,2], Jinyong Leng[1,2], Pu Zhou[1,2,3]**

[1]College of Optoelectronic Science and Engineering, National University of Defense Technology, Changsha 410073, China

[2]Hunan Provincial Collaborative Innovation Center of High Power Fiber Laser, Changsha 410073, China

[3]zhoupu203@163.com

*Email: wujian15203@163.com

**Abstract**

We demonstrate a monolayer graphene based passively Q-switched fiber laser with three-stage amplifiers that can deliver over 80 W average power at 1064 nm. The highest average power achieved is 84.1 W, with pulse energy of 1.67 mJ. To the best of our knowledge, this is the first time for a high power passively Q-switched fiber laser in the 1 μm range reported so far. More importantly, the Q-switched fiber laser operates stably during a week few-hours-per-a-day tests, which proves the stability and practical application value of graphene in high power pulsed fiber lasers.

## 1. Introduction

High power Q-switched fiber lasers are in great demand in material processing, medicine and fundamental research [1-4]. A key approach to generate high average powers is to adopt the master oscillator power amplifier (MOPA) architecture, which usually consists of a seed laser and several stages of power amplifiers. The seed laser provides low power laser emission with stable and low noise output, while the power amplifiers provide power amplification. Up to now, the most powerful pulsed fiber lasers have employed MOPA setup [5-16]. There are some reports on actively Q-switched fiber lasers exploiting MOPA architecture. In the 1 μm range, a Q-switched fiber laser with 62 W average power and 6.2 mJ pulse energy was reported in 2012[15]; another Q-switched fiber laser with 121W average power and 2.4 mJ pulse energy was presented in 2013[16]; a Q-switched fiber laser with pulse energy of 11 mJ[9], and a Q-switched fiber laser with 230 W average output power[12] were demonstrated in 2013, and another 300 W average power Q-switched fiber laser in 2014[13]. In the 2 μm range, a single-frequency actively Q-switched fiber laser with 4.4 W average output power and 220 μJ pulse energy was proposed in 2011[11]; a nanosecond Q-switched fiber laser with 50 W average output, ~1 mJ pulse energy and ~1kW peak power was reported in 2014[14]; and a Q-switched fiber laser that could deliver 110 W average power was fabricated in 2015[10].

Compared with actively Q-switched fiber lasers, passively Q-switched ones, which usually adopt saturable absorbers [17-24], have the advantages of compactness, simple setup and low cost. However, although passively Q-switched fiber lasers have been intensively studied in recent years, there are few reports about high power passively Q-switched fiber lasers except a Q-switched Tm-doped fiber laser based on graphene oxide (GO) presented by Junqing Zhao et al., which delivers ~73 W average power at 1950 nm [1]. Although ytterbium-doped fiber amplifiers (YDFAs) have been proven to be capable of providing high single-pass gain, very high optical-to-optical efficiencies in pulsed laser system, no high power passively Q-switched fiber laser in the 1 μm has ever been reported so far.

In this paper, we build a passively Q-switched high power ytterbium-doped fiber laser based on monolayer graphene that can deliver 84.1 W average power. The pulse operation and stability of the seed laser as well as the performance of the amplifiers are investigated in detail. To the best of our knowledge, this is the first time for a high power passively Q-switched fiber laser reported in the 1 μm range.

## 2. Material Characterization and Experimental Setup

### 2.1 Material Characterization

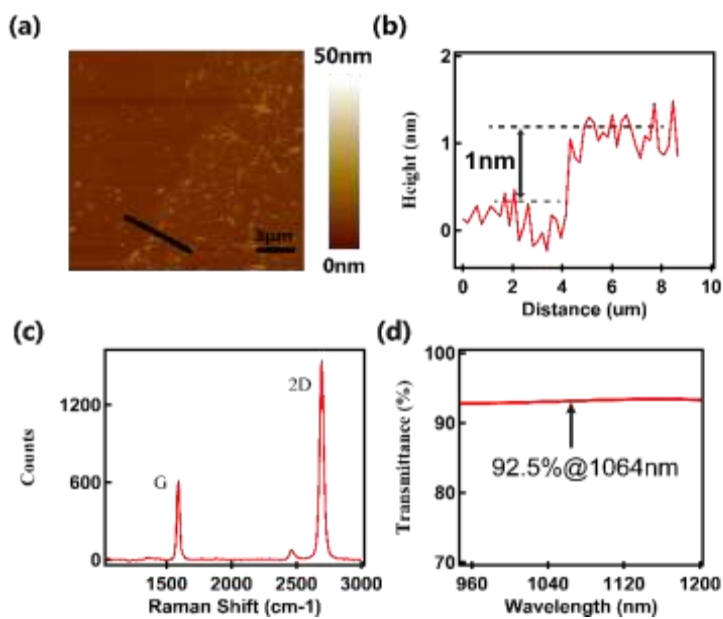

**Figure 1.** Material characterization: (a) AFM image and (b) surface profile of graphene sample (c) Raman shift and (d) transmittance spectrum of graphene sample

The monolayer graphene adopted in our work is bonded on the end face of a fiber patch cord, which is prepared by a similar approach as previous works [25]. The graphene sample is characterized by an atomic force microscopy (AFM), with which the roughness and thickness of sample is measured at the size of 15 μm × 15 μm. The AFM image is shown in Figure 1(a). The surface profile along the black line in Figure 1(a) is plotted in Figure 1(b) and the height difference between the graphene sample and the substrate is about 1 nm, which means the graphene sample is single layer.

Raman scattering measurements are carried out using a commercial confocal microscope Raman system, excited by 473 nm laser. A 50× objective is used for both laser excitation and Raman scattering collection. Figure 1(c) is the Raman shift spectrum of the sample. The prominent G and 2D peaks are located at 1593.3 cm$^{-1}$ and 2693.7 cm$^{-1}$, and the corresponding full width half magnitude (FWHM) are 24 and 37 cm$^{-1}$, respectively. The ratio of G/2D is about 0.39. Thus, the AFM data and Raman spectroscopy can clearly distinguish the monolayer graphene.

At the meantime, the broadband spectrum transmission property of the graphene samples is characterized. A self-constructed supercontinuum source emitting from ~600 nm to ~2000 nm is used to measure the transmission spectrum. As is shown in Figure 1(d), the transmittance is around 92% ~ 93% from 950 nm to 1200 nm and the transmittance at our working wavelength (i.e. 1064 nm) is about 92.5%.

**2.2 Experimental Setup**

The MOPA-based passively Q-switched fiber laser system configuration is depicted in Fig 2, which is comprised of a seed laser followed by multi-stage amplifiers.

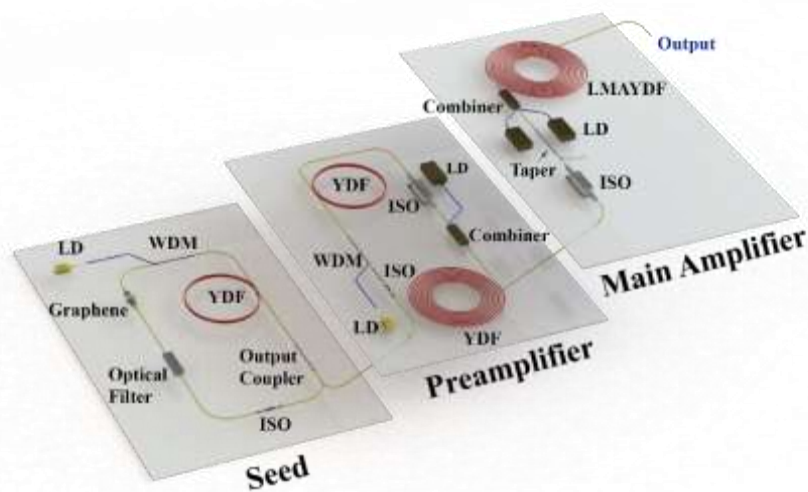

**Figure 2.** Experimental setup of the passively Q-switched fiber laser with 3-stage amplifiers

The seed laser consists of the following components: a single mode laser diode (LD) centered at 976 nm to provide pump power, a wavelength division multiplexer (WDM, 976/1064 nm) to couple the pump light into the ring cavity, a piece of 2.5 meters long ytterbium-doped fiber (YDF, 6/125 μm, NA=0.22) as gain medium, whose absorption is 12 dB/m at 976 nm, an optical filter centered at 1064 nm to select the oscillation wavelength, a 30/70 coupler as output coupler, an isolator (ISO) to ensure unidirectional propagation, and the monolayer graphene sample to serve as saturable absorber. The total cavity length is ~10 m.

The Q-switched pulses are then delivered into the amplifiers. An ISO is inserted between the seed laser and the amplifiers to avoid any reflected light that may induce any instability or damage.

The preamplifier includes two-stage amplifiers. The first-stage amplifier uses a 976/1064 nm WDM to launch the 976 nm pump light into a piece of 2.5 meters long YDF, which is the same as the gain fiber used in the seed laser. The second stage amplifier adopts a piece of 3 meters long double-clad YDF (10/125 μm, core NA= 0.09) whose cladding absorption coefficient is ~6 dB/m at 976 nm , and a 10 W multimode LD centered at 976 nm is used as pump source. The pump light is coupled into the active fiber through a (1+1)×1 combiner. The third stage, i.e. the main amplifier, uses a piece of 3 meters long large mode area (LMA) double-clad YDF (20/130 μm, core NA= 0.09), the cladding absorption of which is ~3.1 dB/m at 915 nm, pumped by two 50 W multimode LDs centered at 976 nm via a (2+1)×1 combiner. There is an isolator between each stage amplifier to block possible counter-propagating light from the subsequent amplifier. A 99/1 taper is inserted before the main amplifier to monitor the backward light in case strong stimulated Raman scattering (SRS) occurs. The LD used in the second- and third-stage as well as the LMA YDF are water-cooled to 20 ℃ to avoid thermal damage.

## 3. Results and discussions

Based on the above-mentioned experimental setup, the overall properties of the passively Q-switched fiber laser with MOPA configuration are investigated. The pulses are measured and recorded by a 5 GHz photodetector together with a 2 GHz oscilloscope. An optical spectrum analyzer (OSA) is used to monitor the output spectrum of the laser system.

**3.1 Performance of the seed laser**

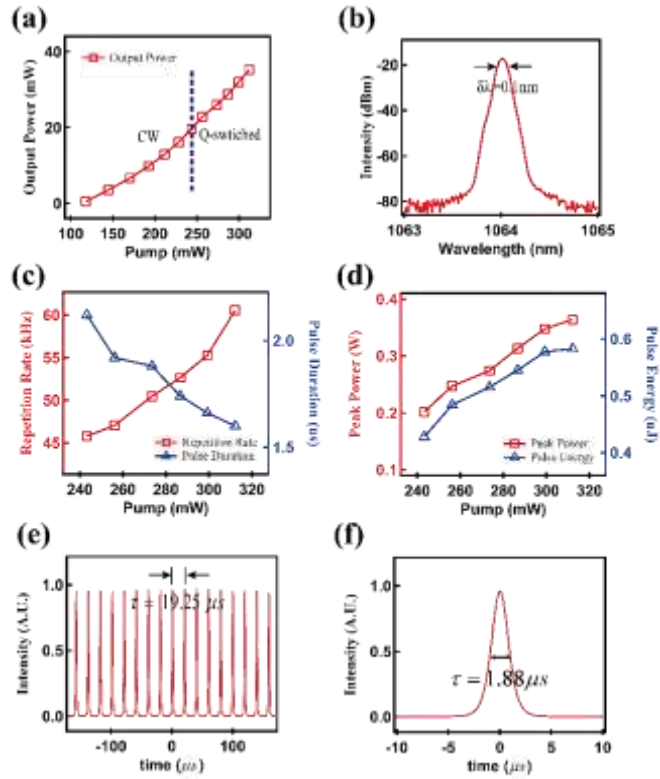

**Figure 3**. Performance of the seed laser: (a) output power against pump power (b) emission spectrum (c) repetition rate and pulse duration and (d) pulse energy and peak power as a function of pump power (e) pulse train and (d) profile of a single pulse at the pump power of 273 mW.

The laser system achieves self-starting Q-switched operation by increasing the pump power over a certain value. Fig 3(a) presents the variation of the output power of the seed laser as a function of the pump power. It starts to lase at the pump power of 118 mW, and goes into Q-switched regime when the pump power exceeds 243.3 mW. The maximum output power reaches 35.3 mW at the pump of 312.2 mW. The overall efficiency is relatively low mainly owing to the insertion loss brought by the graphene and optical filter. Fig 3(b) is the output spectrum of the laser. Since the spectrum is mainly determined by the optical filter, thus no obvious linewidth broaden is observed while increasing the pump power. The linewidth is ~0.1 nm centering at 1064 nm at the maximum pump power.

We also investigated the evolution of Q-switched pulses as a function of the pump power. Fig 3(c) gives the variation of the pulse repetition rate and the pulse duration against the pump power. The repetition rate varies from 45.79 kHz to 60.54 kHz when the pump power increases from 243.3 mW to 312.2 mW, while the pulse duration decreases from 2.12 μs to 1.6 μs simultaneously, which is the typical characteristics of Q-switched operation. Higher repetition rate and shorter pulse duration can be expected if higher pump power is provided. Fig 3(d) depicts the pulse peak power and pulse energy varying with the pump power, both of which increase monotonously: the former of which increases from 0.20 W to 0.36 W, and the latter of which increases from 0.43 μJ to 0.58 μJ.

We fix the pump power at 273 mW, where the output power is 26.01 mW, to make further power amplification, the time interval of adjacent pulses and pulse duration of which are 19.25 μs and 1.88 μs, as shown in Fig 3(e) and (f), respectively.

In order to test the stability of our laser system, we conduct a week of few-hours-per-a-day tests of the seed laser, and record the time interval of adjacent pulses, the pulse duration as well as the output power at the pump of 273 mW. The pulse-pulse interval and pulse duration only experience insignificant change (~0.8% variation of the pulse-pulse interval and ~2% variation of the pulse duration from the average) during the whole tests, as shown in Fig 4(a), which indicates the pulse operation is very stable and can endure long-term tests. Fig 4(b) shows the variation of output power at the pump of 273 mW during a week's per-day tests. The variation is slight (less than 3% variation around the average output power), which further confirms the stability of the seed laser. These results reveal that graphene is an ideal candidate to function as highly stable saturable absorbers in pulsed fiber lasers.

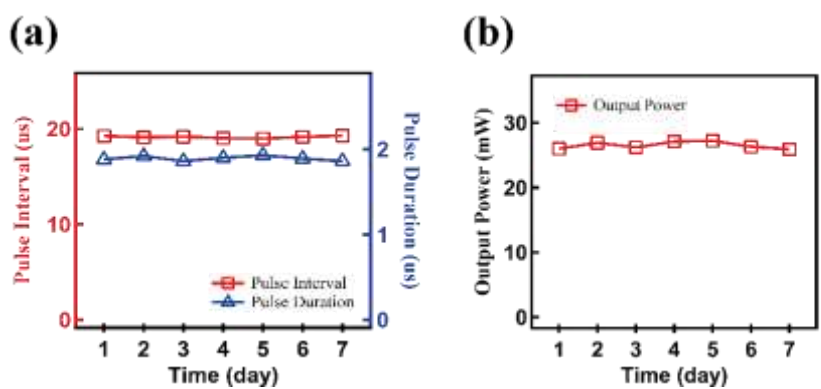

**Figure 4.** Stability of the seed laser: (a) variation of the pulse-pulse interval and pulse duration (b) stability of the output power

**3.2 Performance of the amplifier**

The preamplifier consists of the first two stage amplifiers. The output power is amplified from 26.01 mW to 236.9 mW by the first stage amplifier, after which it is further boosted to 3.95 W by the second stage amplifier. No amplified spontaneous emission (ASE) or SRS is observed under this situation. Then the output of the preamplifier is delivered into the main amplifier to make further power amplification.

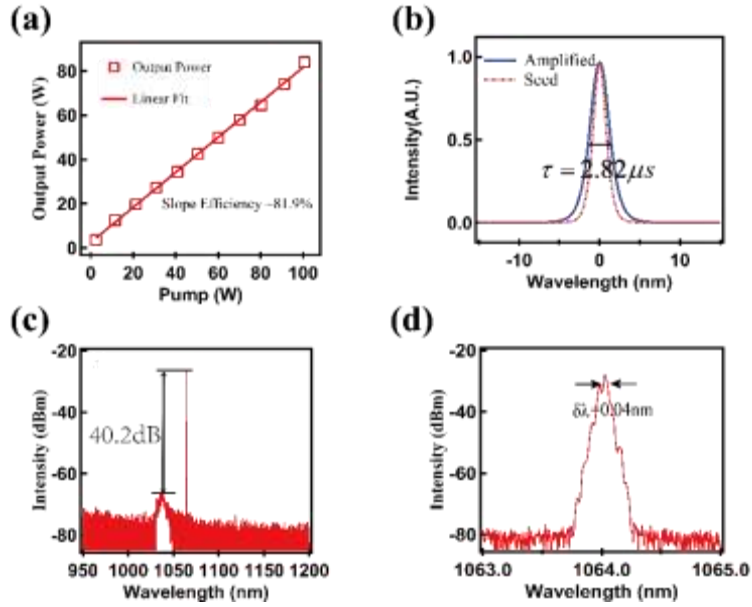

**Figure 5.** Performance of the main amplifier: (a) output power versus pump power (b) profile of the pulse before and after amplification (c) output spectrum in a large wavelength range and (d) fine output spectrum at the pump power of 100.65 W

The performance of the main amplifier is studied in detail, the results of which are depicted in Fig 5. Fig 5(a) presents the output power against the pump power launched into the main amplifier with linear fit. The output power increases almost linearly with a slope efficiency of 81.9 %, with the maximum output power of 84.1 W at the pump of 100.65 W. The profile of a single pulse at the maximum output versus the pulse of the seed laser are depicted in Fig 5(b). The pulse duration increases to 2.82 μs, while the repetition rate remains the same, with peak power of 590.85 W and pulse energy of 1.67 mJ. The spectrum at the maximum pump power spanning from 950 nm to 1200 nm is shown in Fig 5(c). The difference value between the signal amplitude and the maximum ASE intensity can reach 40.2 dB and no parasitic lasing or SRS is observed, which indicates it has potential for further power scaling. The fine spectrum is shown in Fig 5(d), and the linewidth is narrowed to ~0.04 nm.

## 4. Conclusion

In conclusion, we propose and demonstrate a passively Q-switched fiber laser with 84.1 W average output power emitting at 1064 nm based on monolayer graphene. The peak power reaches 590.85 W with pulse energy of 1.67 mJ, and the repetition rate and pulse duration are 50.47 kHz and 2.82 μs, respectively. The slope efficiency of the main amplifier is about 81.9%, and the ASE suppression is about 40.2 dB. No parasitic lasing or SRS are observed in the MOPA. Our work proves the availability and practicality value of graphene saturable absorber in highly stable, high power pulsed fiber lasers. To the best of our knowledge, this is the first time for a passively Q-switched high power fiber laser reported so far in the 1 μm range. The output power is limited by the pump power. Further increase of the output power can be expected by increasing the pump power of the amplifier and adding amplifier stages.